\documentstyle[
				aps,
				twocolumn,
				epsfig,
				]
			{revtex}

\newcommand{\dif}[2]{{\mathrm{#1}}{#2}}
\newcommand{\ra}{\mbox{$\cal{A}$}}
\newcommand{\rb}{\mbox{$\cal{B}$}}
\newcommand{\rl}{\mbox{$\cal{L}$}}

\begin{document}
\twocolumn[
\hsize\textwidth\columnwidth\hsize\csname @twocolumnfalse\endcsname     
\title{Correlated patterns in non-monotonic graded-response 
perceptrons}
\author{D. Boll\'e and T.~Verbeiren}
\address{ Instituut voor Theoretische Fysica, 
Katholieke Universiteit Leuven, B-3001 Leuven, Belgium.}

\maketitle

\begin{abstract} The optimal capacity of graded-response perceptrons
storing biased and  spatially correlated patterns with non-monotonic
input-output relations is studied. It is shown that only the structure 
of
the output patterns is important for the overall performance of the
perceptrons. 
\end{abstract} 
\pacs{PACS numbers: 64.60.Cn, 87.10.+e, 02.50.-r} 
\vskip 3mm ]

%%%%%%%%%%%%%%%%%%%%%%%%%%%%%%%%%%%%%%%%
\section{Introduction}						
%%%%%%%%%%%%%%%%%%%%%%%%%%%%%%%%%%%%%%%% 
Graded-response perceptrons
have been studied intensively in the past years 
(\hspace{-1mm}\mbox{\cite{BE99}} and references therein). 
In particular,  it is found that
for non-monotonic input-output relations interesting retrieval
properties
are obtained such as an improvement of the optimal capacity 
(see, e.g.,\cite{BE96}).

The studies mentioned above concern patterns that are chosen to be 
independent identically distributed random variables with respect to
the sites and the patterns. However, in  practical applications one has
to consider sets of data with internal  structure. While the effects of
bias and correlations on the optimal capacity have been studied  before
for monotonic input-output relations (\hspace{-1mm}\mbox{\cite{G}}- 
\hspace{-3mm}\mbox{\cite{WSKG}} and 
references therein) it is not yet reported on for non-monotonic
ones. This is the purpose of this brief report.

In section \ref{replica}, we write down the final results of a 
Gardner analysis \cite{G} of the capacity problem.  In section
\ref{numerical}, we study these results numerically for some specific  
input-output relations. The influence of spatial input correlations
on the optimal capacity is determined as a function of the
correlation strength. Concerning bias, both the effect of input and
output bias are analysed. Some concluding remarks are presented in 
section  \ref{conclusions}.

%%%%%%%%%%%%%%%%%%%%%%%%%%%%%%%%%%%%%%%%%%%%
\section{Replica analysis}					
\label{replica}
%%%%%%%%%%%%%%%%%%%%%%%%%%%%%%%%%%%%%%%%%%%%%
The graded--response perceptron  maps a
collection of input patterns $\{\xi_i^\mu; 1\leq i\leq N\}$, $1\leq
\mu\leq p=\alpha N$, with $\alpha$ the capacity, onto a corresponding   
set of outputs $\zeta^\mu$ via
\begin{eqnarray}
        \zeta^\mu =  g\left(h^\mu \right) \, , 	 
	\qquad
        h^\mu = \frac{1}{\sqrt N} \sum_j J_j \xi_j^\mu \,.
        \label{eq:2}
\end{eqnarray}
Here $g$ is the input--output relation of the perceptron.  
In (\ref{eq:2}) 
$h^\mu$ is the local field generated by the inputs. 
The $J_j$ are the couplings of the 
perceptron architecture. We focus our attention on general  
input patterns specified by
\begin{equation}
 	\langle \xi_i^\mu\rangle = m 
 \qquad
  	\langle \xi_i^\mu \xi_j^\nu \rangle=\delta_{\mu\nu} C_{ij} \, .
  	\label{eq:3}
\end{equation}
The matrix $C$ formed by the elements $C_{ij}$ is taken to be symmetric
 and 
positive. In the sequel we specifically consider correlations with $m=0$
and general $C$ and correlations with $m \neq 0$ and
$C_{ij}=\delta_{ij}(v + m^2)$. The latter will be called biased patterns
and $v$ is the variance of the input distribution.
 
We allow for a limited output precision in the mapping
(\ref{eq:2}). In other words the output that results when the input 
layer is in the state $\{\xi_i^\mu\}$ is accepted if
\begin{equation}
        g(h^\mu \pm \kappa) \in I_{\rm out}(\zeta^\mu,\epsilon)
                \equiv[\zeta^\mu-\epsilon,\zeta^\mu+\epsilon]
        \label{eq:4}
\end{equation}
where $\epsilon$ denotes the allowed output--error tolerance and
$\kappa$ the required input stability. In order 
to compute the available Gardner--volume \cite{G} in $J$--space 
satisfying 
(\ref{eq:4}) we rewrite the latter as a condition on the local fields 
\begin{equation}
    h^\mu \in I^\mu \equiv \{ x ; g(x \pm \kappa) \in                   
                     I_{\text{out}}(\zeta^{\mu},\epsilon) \} 
    \label{eq:5}
\end{equation} 
where, in general 
\begin{equation}
       I^\mu=\cup_{j=1}^{r^\mu} I^\mu_j = \cup_{j=1}^{r^\mu}            
                   [l_j^\mu,u_j^\mu]
       \label{eq:6}
\end{equation}
form a collection of intervals, not necessarily simply connected, with  
 $l_j^\mu$, $u_j^\mu$ the lower and upper bounds of the $j$-th 
subinterval and $r^\mu$ the number of subintervals defined by the 
pattern $\zeta^\mu$. We remark that for monotonic input-output 
relations, $r^\mu=1$. 
Following the standard Gardner analysis \cite{G} we use the replica 
technique to calculate $v=\lim_{N \to \infty} N^{-1} \left \langle \!  
\left \langle{\ln V} \right \rangle \!  \right \rangle$ with 
$V$ the fractional volume in $J$-space with spherical normalization and 
$\left \langle \!  \left \langle{\cdots}\right \rangle \!  \right 
\rangle$ the average over the statistics of inputs and outputs.

The order parameters occuring in this calculation for correlated 
patterns 
with $m=0$ are \cite{M,TL} 
\begin{eqnarray}
     q_{\lambda \lambda'} &= & \frac{1}{N} \sum_{j,j'} 
                C_{jj'} \, J_j^\lambda J_{j'}^{\lambda'} \, , 
	   \quad \lambda <\lambda' \, ,
		     \\
     Q_\lambda  &=& \frac{1}{N} \sum_{j,j'} 
                C_{jj'} \, J_j^\lambda J_{j'}^\lambda \, , 
		\quad               
               \lambda,\lambda'=1,\ldots ,n
	\label{eq:orderu}
\end{eqnarray}
with $n$ the number of replicas.
Since the set of general fixed-point equations leading to the optimal 
capacity $\alpha_c$ (obtained when $V=0$) in the replica-symmetric (RS)
approximation  has been discussed already in \cite{M,TL} (for a 
simple perceptron with the sign-function as input-output relation)
we do not write out explicitly the 
analogous formula for the graded-response perceptron with correlated 
input and binary output but non-monotonic input-output relations.  For 
technical
reasons the latter are taken to be odd in the field. We just mention 
that the essential difference is a splitting of the integrations in 
regions of the form $[(u_{j-1}+l_j)/2,l_j]$ and 
$[u_{j},(u_j+l_{j+1})/2]$ corresponding to the collection of intervals 
(\ref{eq:6}) (compare eq.~(8) in \cite{BE96}). No closed form for 
$\alpha_c$ is possible and  the solution of these fixed-point equations 
is rather tedious. In the next section we present some numerical 
results for exponentially decaying spatial correlations. 

For biased patterns ($m \neq 0$ and $C_{ij}=\delta_{ij}(v + m^2)$) the 
order 
parameters read
\begin{eqnarray}
     q_{\lambda \lambda'} &=& \frac{1}{N} \sum_{j} 
                  J_j^\lambda J_{j}^{\lambda'} \, ,
		  \quad \lambda < \lambda',
		          \\
 M_\lambda  &=& \frac{1}{\sqrt{N}} \sum_{j} J_j^\lambda\, ,  \quad
    \lambda,\lambda'=1,\ldots ,n \,.
	\label{eq:orderb}
\end{eqnarray}
Since in this case a closed form for $\alpha_c$ is possible and its
structure  is  
interesting for analysing the effects of the bias $m$ we write it down 
explicitly in a first-step replica-symmetry breaking approximation
(RSB1). Applying the Parisi scheme \cite{BE96,MPV} 
we find 
%after a standard but tedious calculation
\begin{equation}
   \alpha^{RSB1}_c=\min_{P,q_0} \max_{M}  
      \frac{-\ln(1+P(1-q_0))-\displaystyle{\frac{P q_0}{1+P(1-q_0)}}}
       {\displaystyle{2\left \langle \int      
    \dif{D}{z_0}\ln H(I^\mu,q_0,P,z_0)\right 
					 \rangle_{\zeta^\mu}}}
        \label{eq:alphac}
\end{equation}
with, dropping the index $\mu$ in the sequel
\begin{eqnarray}                                                        
        H&=&\sum_{j=1}^{r}\rl\left ( -\rb(l_j),
          -\frac{1}{2}[\rb(u_{j-1})+\rb(l_j)],y(l_j) \right )
                   \nonumber \\ 
          & +&\rl\left ( -\frac{1}{2}[                          
          	\rb(u_{j})+\rb(l_{j+1}],-\rb(u_j),y(u_j) \right )
		    \nonumber \\
          &+&\rl\left ( -\rb(u_{j}),-\rb(l_j),0 \right )
\end{eqnarray}
where $u_0=-\infty$, $l_{r+1}=+\infty$,
\begin{eqnarray}
    \rl(a,b,c) &=& \int_a^b \dif{D}{z_1} \exp\left ( -\frac{1}{2}P c^2  
                    \right )\, ,\\
        \rb(x) &=& \frac{\ra(x)+ z_0 \sqrt{q_0}}{\sqrt{1-q_0}}\, , \quad
	\ra(x) = {x-mM \over \sqrt{v}} \, ,\\
       y(x) &=& \ra(x)+z_0 \sqrt{q_0} +z_1 \sqrt{1-q_0} \, ,
\end{eqnarray}
with $\dif{D}{z}=dz (2\pi)^{-1/2} \exp(-z^2/2)$ the Gaussian measure. 
We 
remark that for zero bias  we find back the results given in 
\cite{BE96}.

%%%%%%%%%%%%%%%%%%%%%%%%%%%%%%%%%%%%%%%%%%%%%%%%%%%%%%%%%
\section{Results}		\label{numerical}
%%%%%%%%%%%%%%%%%%%%%%%%%%%%%%%%%%%%%%%%%%%%%%%%%%%%%%%%%
Two input-output relations are studied for comparison. The piecewise 
linear 
one
\begin{equation}
   g_{\mbox{\tiny{L}}}(x)= \left \{  \begin{array}{l@{\hspace{1cm}}r}
                 {\mathrm{sgn}}(x) & |x| \ge 1/\gamma \\
                  \gamma x &  |x| < 1/\gamma
                  \end{array}
                \right.
       \label{eq:mon}
\end{equation}
as a prototype of a general monotonic function, and the reversed-wedge
\cite{WR,BMZ} 
\begin{equation}
g_{\mbox{\tiny{RW}}}(x)={\mathrm{sgn}}[(x+1/\gamma)x(x-1/\gamma)]
        \label{eq:nonmon} 
\end{equation}
as an example of a non-monotonic one. Here $\gamma$ is called the gain 
parameter. In the numerical analysis we restrict ourselves mostly to 
$\epsilon=0$.  

%%%%%%%%%%%%%%%%%%%%%%%%%%%%%%%%%%%%%
\subsection{Correlated patterns}
%%%%%%%%%%%%%%%%%%%%%%%%%%%%%%%%%%%%%
Following \cite{M}, we study correlations in the  input 
patterns that are positive and fall off with the distance between the 
sites
\begin{equation}
    C_{ij}=\exp\left ( -\frac{|i-j|}{L} \right ) \equiv S^{|i-j|} \ ,
     \label{eq:cor}
\end{equation}
with $L$ a typical length size.  The parameter $S$ is the 
correlation strength inside one input pattern and varies between $0$, 
corresponding to independent sites and $1$ meaning that all 
spins in a pattern are equal.
The spatial structure introduced above induces interesting correlations 
between the couplings. The latter are positively correlated when close
enough and anticorrelated when further apart \cite{M}.
 
\begin{figure}
\centerline{\hbox{
\epsfig{figure=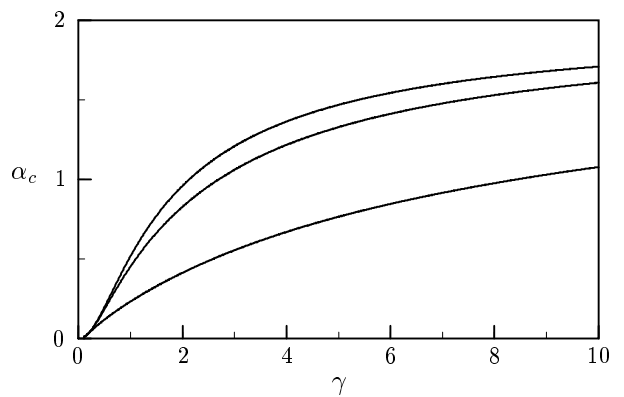,clip=}}}
\caption{The optimal capacity $\alpha_c$ as a function of $\gamma$ for
  $g_{\mbox{\tiny{L}}}$ with correlated inputs.
  From top to bottom $S=0,.5,.9$}
\label{figure1}
\end{figure}

For the sign-function it has been shown \cite{M} that the optimal
capacity, $\alpha_c$, at $\kappa=0$ remains $2$ regardless of the inner 
structure of the inputs. 
Here we analyse in addition $\alpha_c$ as a function of $\gamma$ for    
different correlation strengths $S$ in the case of 
$g_{\mbox{\tiny{L}}}$. 
The results are shown in Fig.~\ref{figure1}.
Increasing $\gamma$
or decreasing $S$, $\alpha_c$ increases. In the limit $\gamma \to
\infty$, $g_{\mbox{\tiny{L}}}$ becomes the sign-function such that  
$\alpha_c$ always approaches $2$ because of the argument above.  

Numerically we have found that changing $S$ can be seen as an 
effective scaling in $\gamma$. Since this scaling behavior can be 
shown also analytically for biased patterns we write it down in 
the next subsection.

\begin{figure}
\centerline{\hbox{
\epsfig{figure=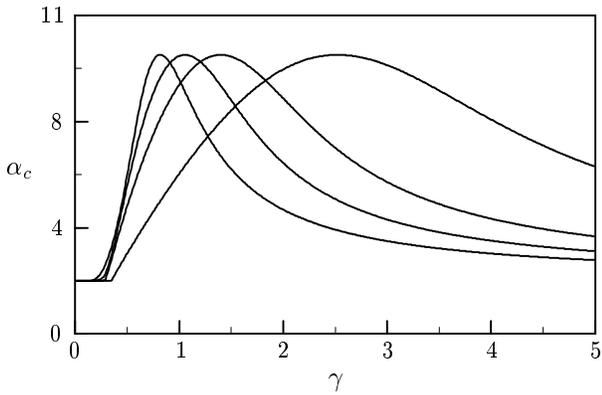,clip=}}}
\caption{The optimal capacity $\alpha_c$ as a function of  $\gamma$ 
for $g_{\mbox{\tiny{RW}}}$  with correlated inputs. 
From left to right $S=0,.2,.7,.9$}
\label{figure2}
\end{figure}

For the non-monotonic input-output relation $g_{\mbox{\tiny{RW}}}$ the  
corresponding results are presented in Fig.~\ref{figure2}.
Several remarks are in order. Technically, the solutions of 
the relevant saddle-point equations are unique for small values of $S$ 
for all $\gamma$ but from $S>S_c=0.55$ onwards there exist 
multiple solutions in a growing interval in $\gamma$. This is due to 
the 
non-monotonicity of the input-output relation. Taking the solution 
giving 
the greatest optimal capacity we find the constant part $\alpha_c=2$ 
(for small $\gamma$ and $S>S_c=0.55$) of the curves in
Fig.~\ref{figure2}. It seems that for these values of $\gamma$, the 
perceptron is not able to benefit from the non-monotonicity of the 
input-output relation due to the fact that the order parameter $Q$ 
remains 
small. 

It is interesting to see that the maximal value of $\alpha_c$ is
the same independent of the correlation strength meaning that also for 
the non-monotonic $g_{\mbox{\tiny{RW}}}$ the inner structure of the 
patterns 
does not play any role in this respect. This precisely amounts
to a scaling of $\gamma$.

%%%%%%%%%%%%%%%%%%%%%%%%%%%%%%%%
\subsection{Biased patterns}
%%%%%%%%%%%%%%%%%%%%%%%%%%%%%%%
In this section we study biased input and output patterns. Their 
probability distribution is chosen to be
\begin{equation}
   \rho(x)=\frac{1+m}{2}\delta(1-x)+\frac{1-m}{2}\delta(1+x)
   \label{eq:distribution}
\end{equation}
where $m$ can be different for input and output, thus defining $m_i$ 
and $m_o$.

We start with some general properties of the perceptrons defined by
eqs.~(\ref{eq:mon}-\ref{eq:nonmon}). Comparing the
results  (\ref{eq:alphac}) with those of            
\cite{BE96,BKM}, we see that in order to obtain the expressions for 
biased patterns it is sufficient to substitute the local field $h$ by 
$(h-m_iM)/\sqrt v$, and to perform an extra maximization over $M$ in the
expressions for patterns without bias. 
This tells us that the order parameter $M$ indicating the bias in       
the couplings, as seen in its definition (\ref{eq:orderb}), shifts the
local field such that the condition (\ref{eq:5}) is optimally satisfied 
and hence that the capacity increases. Furthermore, it naturally
introduces two
cases: $m_iM=0$ and $m_iM \neq 0$. Whenever $m_i$ or $m_0$ are zero
$m_iM=0$. However, $m_iM=0$ does not
necessarily imply that $m_i$ or $m_o$ are zero as we will see
explicitly in the case of the non-monotonic $g_{\mbox{\tiny{RW}}}$. But 
these
points where $m_iM=0$ occur rather exceptionally. 

A closer inspection of the results in section \ref{replica} shows that
the graded-response perceptron satisfies the following analytic scaling 
behavior, given an output distribution:
\begin{eqnarray}
 && \hspace*{-1cm} 
    \alpha_c \left( m_{i,1},m_o;
     \gamma \sqrt{v_1},\epsilon,\frac{\kappa}{\sqrt{v_1}}\right) 
            \nonumber\\ &\equiv&
    \alpha_c \left( m_{i,2},m_o;
     \gamma \sqrt{v_2},\epsilon,\frac{\kappa}{\sqrt{v_2}}\right) 
        	 \label{eq:scaling} 
\end{eqnarray}
where  $m_{i,1}$ and $m_{i,2}$ are two values of the input bias and 
$v_1$
and $v_2$ two corresponding  values of the variance. 
These results are valid for both monotonic and non-monotonic
input-output relations. The new insight is that the output
statistics  is the  
important quantity determining the performance of the perceptron. In
general, increasing the bias in the output results in
a non-decreasing optimal capacity.

Concerning RS stability assured by a negative sign of the
replicon eigenvalue $\lambda_R$ \cite{MPV} we know that for
monotonic non-decreasing input-output relations and unbiased patterns 
the   
following identity holds (compare \cite{BKM})
\begin{eqnarray}
 &&\hspace*{-1cm}
   \mbox{sgn}[-\lambda_R(m_i,m_o=0;\gamma,\epsilon,\kappa=0)]
   \nonumber \\ &=&
     \mbox{sgn} \left[ \frac{\partial}{\partial \gamma} \,
	 \alpha_c(m_i,m_o=0;\gamma,\epsilon,\kappa=0) 
	 \right] \,.
\end{eqnarray}
Together with (\ref{eq:scaling}) this relation tells us that
varying the input bias does not change the breaking behavior for a fixed
$\gamma$. The scaling (\ref{eq:scaling}) also implies  
\begin{eqnarray}
  &&\hspace*{-1cm} 
     \mbox{sgn}[-\lambda_R(m_i,m_o=0;\gamma,\epsilon,\kappa=0)]
   \nonumber \\ &=&
      \mbox{sgn} \left[ \frac{\partial}{\partial m_i} \,
	  \alpha_c(m_i,m_o=0;\gamma,\epsilon,\kappa=0)
             \right]\,.
\end{eqnarray}
For non-monotonic transfer functions we know that replica-symmetry is
unstable \cite{B}.

For the graded-reponse perceptron with monotonic input-output relation 
$g_{\mbox{\tiny{L}}}$ and pattern distribution (\ref{eq:distribution})
we  find the following  additional results  concerning the output bias.
For $M=0$ the solution is stable for all values of the bias in input 
and output, for each $\gamma$. 

\begin{figure}
\centerline{\hbox{
\epsfig{figure=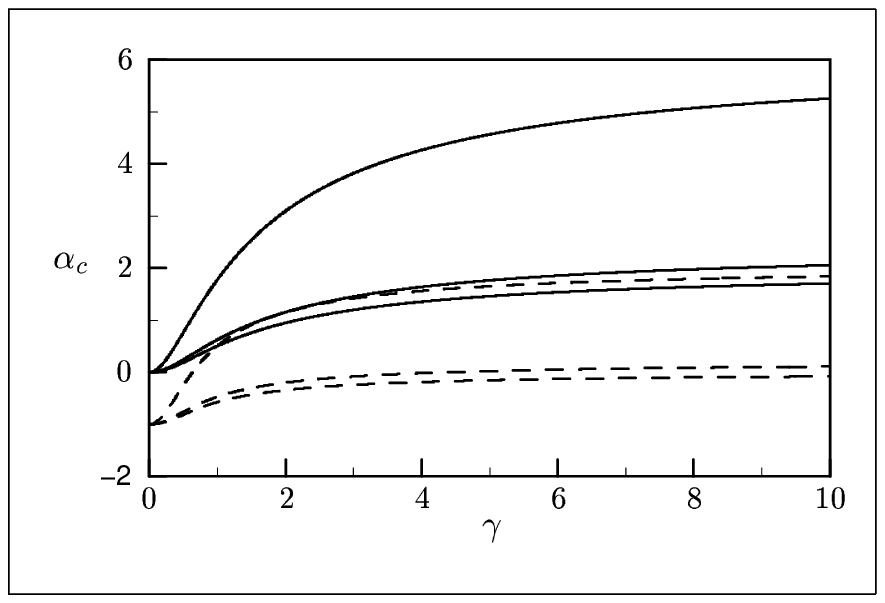,clip=}}}
\caption{The optimal capacity $\alpha_c$ (full curves) and the replicon 
eigenvalue $\lambda_R$ (dashed curves) for $g_{\mbox{\tiny{L}}}$ as a   
function of $\gamma$ with $m_i=.2$ and, from bottom to top 
$m_o=0,.5,.9$.} 
\label{figure3}
\end{figure}

For $M \neq 0$, from a certain value of $m_o$ onwards the RS solution   
becomes unstable for a growing interval of $\gamma$-values. This is
shown in Fig.~\ref{figure3}. 
For these  perceptrons it is known \cite{BE96,BKM} that  
the effect of breaking is small.  Although it grows with 
increasing output-bias, it is seen that for $m_o<.9$, the difference in 
capacity does not exceed $10^{-2}$.
We remark that the maximum capacity as a function of $\gamma$ is 
reached 
for $\gamma \to \infty$ in agreement with the result obtained in 
\cite{G}.

\begin{figure}
\centerline{\hbox{
\epsfig{figure=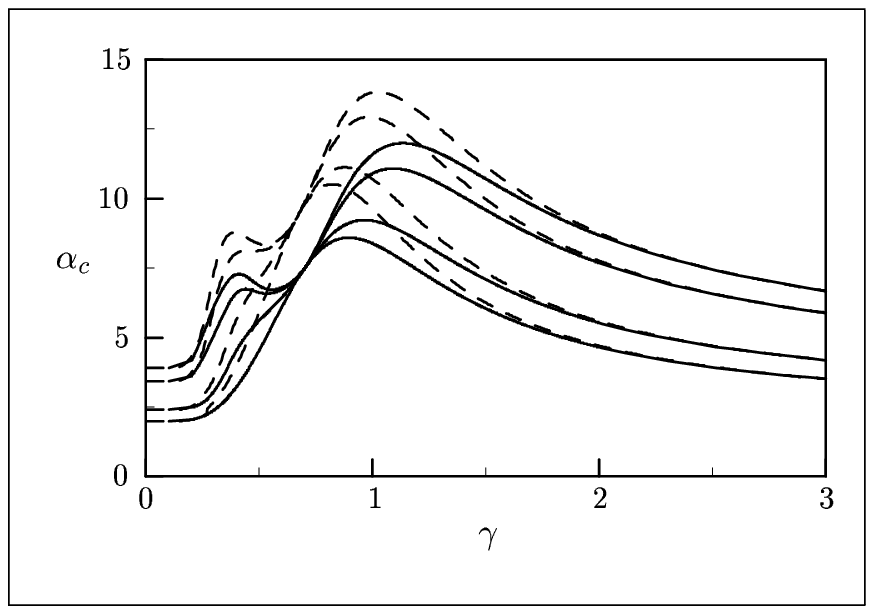,clip=}}}
\caption{The RS (dashed curve) and RSB (full curve) optimal capacity    
$\alpha_c$ of $g_{\mbox{\tiny{RW}}}$ as a function of
$\gamma$  with $m_i=.1$ and, from bottom to top, $m_o=0,.5,.75,.8$.}  
\label{figure4}
\end{figure}

For the non-monotonic transferfunction $g_{\mbox{\tiny{RW}}}$ the 
maximal
$\alpha_c$ is obtained for a finite value of $\gamma$, as shown in 
Fig.~\ref{figure4} implying that there exists an optimal choice for the 
width of the plateaus. This choice depends on the specific parameters of
the pattern distribution.

Compared with the monotonic case the overall difference between the
RS and RSB1 solution is much bigger.
The optimal capacity for the non-monotonic perceptron is always greater
than that of the monotonic one. We note that for $\gamma \to 0$ and 
$\gamma \to \infty$ the optimal capacity of $g_{\mbox{\tiny{RW}}}$
approaches  that of the sign-function.  
We remark that, as in the case of correlated input patterns, the
order-parameter $M$ gives rise to multiple solutions for small values of
$\gamma$. We take the solution with the  highest $\alpha_c$.

A somewhat surprising feature of this perceptron is that a second
maximum develops both in the RS and RSB1-solution as a
function of $\gamma$ for big values of $m_o$ (see Fig.~\ref{figure4}).
Qualitatively speaking the overall behavior of the input-output
relation remains the same within RSB1.  This is the case for all values
of the model parameters we have considered but may be a property of
the binary output distribution (compare \cite{BE96} for a uniform
output). The difference between RS and RSB1 grows with increasing bias. 

Between the two maxima, there is a point where $\alpha_c$ does not 
depend on the output $m_0$. This feature is present both in the RS and
the RSB1-approximation although for a slightly different value of
$\gamma$ in RSB1.  The
underlying reason for this is that the solution of $M$ at these points
is zero and that the input-output relation is odd in the local field,
such that the output statistics does not effect the optimal capacity of
the system.  Since changing $m_i$ can be expressed as rescaling
$\gamma$, the capacity at these points is the 
same for every value of $m_i$ and $m_o$.

We end with the remark that changing the input distribution  
(\ref{eq:distribution}) by varying the place of the delta peaks in the
interval $[0,1]$ shows a similar scaling behavior.

%%%%%%%%%%%%%%%%%%%%%%%%%%%%%%%%%%%%%%%%
\section{Conclusions}		\label{conclusions}
%%%%%%%%%%%%%%%%%%%%%%%%%%%%%%%%%%%%%%
In this brief report we have studied the optimal capacity of
graded-response perceptrons storing biased and spatially correlated
patterns with non-monotonic input-output relations using a  first-step
replica-symmetry breaking analysis.

The most important results are that a change in the optimal capacity 
due to 
bias or correlations 
in the input can be removed by an appropriate scaling of the relevant
parameters defining the graded-response perceptron.  The statistics of
the outputs really determines the performance of the latter.

%---------

\section*{Acknowledgments}		
The authors would like to thank Dr. Geert Vancraeynest for informative
discussions in the initial stages of this work.  They are indebted to
the Fund for Scientific Research Flanders (Belgium) for financial
support.

%---------

\bibliographystyle{prsty}

\begin{references}
\bibitem{BE99} D. Boll\' e and R. Erichsen Jr., Phys. Rev. E {\bf 59}, 
3386 (1999). 
\bibitem{BE96} D. Boll\' e and R. Erichsen Jr., J. Phys. A: Math. Gen. 
{\bf 29}, 2299 (1996).
\bibitem{G} E. Gardner, Europhys. Lett. {\bf 4}, 481 (1987); 
J. Phys. A: Math. Gen. {\bf 21}, 257 (1988).
\bibitem{M} R. Monasson, J. Phys. A: Math. Gen. {\bf 25}, 3701 (1992). 
\bibitem{TL} W. Tarkowski and M. Lewenstein, J. Phys. A: Math. Gen. 
{\bf 26}, 2453 (1993).
\bibitem{WSKG} J.O. Winkel, B. Schottky, U. Krey and F. Gerl, Z. Phys. B
{\bf 101}, 305 (1996).
\bibitem{WR} T.L.H. Watkin and A. Rau, Phys.Rev. A {\bf 45}, 4102
  (1992).
\bibitem{BMZ} G. Boffetta, R. Monasson and R. Zecchina, J. Phys. A: 
Math.
Gen. {\bf 26}, L507 (1993).  
\bibitem{MPV}
M. M\'ezard, G. Parisi and M.A. Virasoro, {\it Spin Glass Theory
and Beyond} (World Scientific, Singapore, 1987)
\bibitem{BKM} D. Boll\' e, R. Kuhn and J. van Mourik, 
J. Phys. A: Math. Gen. {\bf 26}, 3149 (1992).
\bibitem{B}M. Bouten, J. Phys. A: Math. Gen. {\bf 27}, 6021 (1994).

\end{references}

\end{document}